\documentclass[12pt]{article}
\usepackage{latexsym,graphicx,a4,epsfig,psfrag,here}

\newcommand{\bda}{\begin{\displaymath}\begin{array}{rl}}
\newcommand{\eda}{\end{array}\end{displaymath}}
\newcommand{\be}{\begin{equation}}
\newcommand{\ee}{\end{equation}}
\newcommand{\bdm}{\begin{displaymath}}
\newcommand{\edm}{\end{displaymath}}
\newcommand{\bea}{\begin{eqnarray}}
\newcommand{\eea}{\end{eqnarray}}
\newcommand{\no}{\nonumber \\}
\newcommand{\fs}{\,.}
\newcommand{\co}{\,,}
\newcommand{\al}{&\!\!\!}

\newcommand{\g}{\rule{0.5em}{0em}}
\begin{document}

\thispagestyle{empty}

\vspace{3cm}
\begin{center}
{\LARGE\bf {On the precision of the theoretical predictions for $\pi\pi$
    scattering}}

\vspace{0.5cm}
August 22, 2003

\vspace{0.5cm}
I.~Caprini$^a$, G.~Colangelo$^b$,
J.~Gasser$^b$ and
H.~Leutwyler$^b$

\vspace{2em}
\footnotesize{\begin{tabular}{c}
$^a\,$ National Institute of Physics and Nuclear Engineering\\
POB MG 6, Bucharest, R-76900 Romania\\
$^b\,$Institute for Theoretical Physics, University of 
Bern\\
Sidlerstr. 5, CH-3012 Bern, Switzerland
\end{tabular}  }

\vspace{1cm}

\begin{abstract}
In a recent paper, Pel\'aez and Yndur\'ain evaluate some of the low energy
observables of $\pi\pi$ scattering and obtain flat disagreement with our
earlier results. The authors work with unsubtracted dispersion relations,
so that their results are very sensitive to the poorly known high energy
behaviour of the scattering amplitude. They claim that the asymptotic
representation we used is incorrect and propose an alternative one. We
repeat their calculations on the basis of the standard, subtracted
fixed-$t$ dispersion relations, using their asymptotics. The outcome fully
confirms our earlier findings. Moreover, we show that the
Regge parametrization proposed by these authors for the region above 1.4 GeV
violates crossing symmetry: Their ansatz is not consistent with the
behaviour observed at low energies.
\end{abstract}

\end{center}
\newpage
\section{Introduction}
We have demonstrated that the low energy properties of the $\pi\pi$ scattering
amplitude can be predicted to a remarkable degree of accuracy \cite{ACGL,CGL}
(in the following these papers are referred to as ACGL and CGL, respectively).
In our opinion, this work represents a breakthrough in a field that
hitherto was subject to considerable uncertainties. The low energy
properties of the $\pi\pi$ scattering amplitude play a central role in the
analysis of many quantities of physical interest. As an example, we mention
the magnetic moment of the muon, where the Standard Model prediction requires
precise knowledge of the hadronic contributions to vacuum polarization. As
these are dominated by two-pion intermediate states of angular momentum
$\ell=1$, the P-wave $\pi\pi$ phase shift is needed to high accuracy in
order to analyze the data in a reliable manner \cite{Arkadyfest,paper on form
  factor}.

Our dispersive analysis, which is based on the Roy equations \cite{Roy},
was confirmed\footnote{This paper also compares
our predictions for the values of the two subtraction constants with some of
the $\pi\pi$ phase shift analyses and with the new $K_{e_4}$ data obtained by
the E856 collaboration at Brookhaven. While the result
obtained in ref.~\cite{Descotes} for $a_0^0$ is consistent with the theoretical
prediction, the one for the combination $2a_0^0-5a_0^2$ deviates from the
value predicted in CGL by 1 $\sigma$. Further work on this issue is reported
in ref.~\cite{Maiorov Patarakin}.} in ref.~\cite{Descotes}. 
In a recent paper, however, Pel\'aez 
and Yndur\'ain \cite{PY} claim that this analysis is deficient, because the
representation we are using to describe the behaviour of the imaginary
parts above 1.42 GeV is ``irrealistic''. They propose an alternative
representation, evaluate a few quantities of physical interest on that
basis and obtain flat disagreement with our results. They conclude that our
solution to the constraints imposed by analyticity, unitarity and chiral
symmetry is ``spurious''. In the following, we refer to this paper as PY and
show that this claim and others contained therein are incorrect.

As a first step, we briefly outline our framework. The fixed-$t$ dispersion
relations of Roy represent the real parts of the scattering amplitude in
terms of the $s$-channel imaginary parts and two subtraction constants,
which can be identified with the two S-wave scattering lengths, $a_0^0,
a_0^2$. The Roy equations represent the partial wave projections of these
dispersion relations. Since the partial wave expansion of the imaginary
parts converges in the large Lehman-Martin ellipse, it follows from first
principles that the Roy equations hold for $-4M_\pi^2< s < 60 M_\pi^2$,
i.e.~up to a centre of mass energy of $1.08\,\mbox{GeV}$. We use these
equations to determine the phases of the S- and P-waves on the interval
$2M_\pi< \sqrt{s}<0.8\,\mbox{GeV}$. The calculation treats the imaginary
parts above 0.8 GeV as well as the two subtraction constants as external
input.

As demonstrated in ACGL, the two subtraction constants play the key role
in the low energy analysis. The central observation in CGL is
that the values of these two constants can be predicted on the basis of chiral
symmetry. Weinberg's low energy theorem
\cite{Weinberg 1966} states that, to leading order in the expansion in powers
of $m_u$ and $m_d$, the scattering lengths $a_0^0$ and $a_0^2$ are determined
by the pion decay constant. The corrections are known up to and including
next-to-next-to-leading order \cite{BCEGS}. In CGL, we have performed a new
determination of the relevant effective coupling constants, thereby
obtained sharp predictions for $a_0^0,a_0^2$ and then demonstrated
that the Roy equations pin down the $\pi\pi$ scattering
amplitude throughout the low energy region, to within very small uncertainties.

The paper is organized as follows. We first discuss the difference between PY
and CGL concerning the input used for the imaginary parts
in the region above 1.42 GeV. In sections
\ref{sec:low energy theorem}-\ref{sec:threshold parameters}, we then repeat
the calculations reported in CGL for the input advocated by
Pel\'aez and Yndur\'ain, who did not 
perform such an analysis, but claim that the results are sensitive to the
input used in the asymptotic region. As we will demonstrate explicitly, this
is not the case. We turn to the calculations they did perform only in
the second part of the paper, where we show that their
Regge representation cannot be right because it violates crossing
symmetry. Section 10 contains a summary of the present article as well as our
conclusions. 

\section{Asymptotics}
\label{sec:asymptotics}
According to PY, the input used for the imaginary parts above 1.42 GeV
plays an important role in our analysis. This contradicts the findings in
ACGL, where we demonstrated explicitly that the behaviour at those energies
is not essential, because the integrals occurring in the Roy equations
converge rapidly. In particular, our explicit estimates for the sensitivity
of the threshold parameters to the input used at and above 0.8 GeV (see
table 4, column $\Delta_1$ in ACGL) imply that the uncertainties from this
source are very small. In view of this, it is difficult to understand the
claim of PY that our solutions are ``distorted'' because the input used for
$\sqrt{s}>1.42\,\mbox{GeV}$ is ``irrealistic''.

Admittedly, however, we did not perform a thorough study of the imaginary
parts for energies above 1.42 GeV -- for brevity we refer to this range as
the asymptotic region. In the interval from 1.42 to 2 GeV, we relied on
phenomenology, while above 2 GeV, we used a Regge representation based on
the work of Pennington and Protopopescu \cite{Pennington
Protopopescu,Pennington}. In particular, we used their results for the
residue of the Regge pole with the quantum numbers of the $\rho$ meson,
also with regard to the uncertainties to be attached to this contribution,
and invoked a sum rule that follows from crossing symmetry to estimate the
magnitude of the Pomeron term.

According to Pel\'aez and Yndur\'ain, phenomenology cannot be trusted up to
2 GeV. The authors construct what they refer to as an ``orthodox'' Regge fit
and then assume that this fit adequately approximates the imaginary parts
down to a centre of mass energy of 1.42 GeV. For ease of comparison, the
Regge representation of PY is described in appendix \ref{sec:Regge PY}. It
differs significantly from ours. Moreover, in the region below 2 GeV, it
differs from the phenomenological input we used.  Although we attached
considerable uncertainties to the input of our calculation, these do not
cover the asymptotic representation proposed in PY.

Unfortunately, the authors do not offer a critical discussion of their
representation, which looks similar to the Regge fit proposed by Rarita et
al.~\cite{Rarita} in 1968, but the parameters are assigned different values
and a comparison is not made. For a review of the current knowledge about the
structure of the Pomeron, we refer to \cite{Landshoff}. Recent thorough
analyses of different classes of parametrizations of the asymptotic
amplitudes and of the corresponding fits to the large body of available data
are described in \cite{PDG,Cudell}. These indicate that the leading terms can 
be determined rather well by applying factorization to the experimentally well
explored $N N$ and $\pi N$ scattering amplitudes, but the non-leading
contributions become more and
more important as the energy is lowered (see, e.g., \cite{Cudell} for a
critical discussion of the range of applicability of different asymptotic
formulae).  We do not consider it plausible that the asymptotic
representation of PY can be trusted to the precision claimed in that paper,
where the uncertainties in the contributions from the region above 1.42 GeV
are estimated at 10 to 15 \%.

In the following, however, we take the asymptotic representation proposed
in PY at face value. More precisely, we (i) replace our Regge
parametrization by this one and (ii) set $s_0=(0.8\,\mbox{GeV})^2$, $s_2=
(1.42\,\mbox{GeV})^2$. All other elements of the calculation are taken over
from CGL without any change, so that we can study the sensitivity of the
result to the asymptotics.  We solve the Roy equations between threshold
and $s_0$, rely on phenomenological information about the imaginary parts
on the interval from $s_0$ to $s_2$ and use the Regge representation of PY
above that energy.

In PY, a further contribution is added, to account for the enhancement in
the $I=1$ imaginary part associated with the $\rho(1450)$. The
corresponding contributions to the various observables considered in PY are
explicitly listed there. In all cases, these are smaller than our estimates
for the uncertainties to be attached to our results. In the following, we
drop this term to simplify the calculations. Note also that in PY, a
parametrization for the D- and F-waves is used that is somewhat different
from those we rely on, which are taken from refs.~\cite{hyams,bugg}.

\section{Low energy theorem for {\boldmath $a_0^0$ and $a_0^2$}}
\label{sec:low energy theorem}
\begin{table}[thb]
\begin{center}
\begin{tabular}{|c||r|r|}
\hline
&\multicolumn{1}{|c|}{CGL}&\multicolumn{1}{|c|}{PY}\\
\hline\hspace{-0.3em}\rule{0em}{1em}
$a_0^0$&$ 0.220\pm .005$\rule{0.5em}{0em}&
$\rule{0.3em}{0em}0.221$\rule{0.5em}{0em}
\\
\hline\hspace{-0.3em}\rule{0em}{1em}
$a_0^2$&$-0.0444\pm .0010$&$-0.0443$
\\
\hline\hspace{-0.3em}\rule{0em}{1em}
$2\,a_0^0-5\,a_0^2$&$0.663\pm .007$\rule{0.5em}{0em}&$0.663$\rule{0.5em}{0em}
\\\hline
\end{tabular}
\end{center}
\caption{\label{tab:S-wave scattering lengths}
S-wave scattering lengths. The numbers in the first column are taken from
CGL. Those in the second column are obtained by replacing the
asymptotics used there with the one proposed in PY.}
\end{table}
The low energy structure is controlled by the two
subtraction constants. The main question to ask, therefore, is whether the
change in the asymptotics proposed in PY affects the predictions for these
two constants. In principle, it does, because
some of the corrections to Weinberg's low energy theorem \cite{Weinberg 1966}
involve integrals
over the imaginary parts of the scattering amplitude that extend to
infinity. As documented in table 1
of CGL, the uncertainties in the result for the $S$-wave
scattering lengths are dominated by those in the effective coupling
constants. The noise in the input used at and above 0.8 GeV 
affects the values of $a_0^0$ and $a_0^2$ only at the level of half a
percent. 

As mentioned above, however, our estimates for the uncertainties in the
asymptotic part of the input do not cover the modification proposed in
PY. To remain on firm grounds, we have repeated the calculation
described in CGL, using as input above 1.42 GeV the parametrization proposed
in PY. We have also reexamined the dispersive evaluation of the scalar radius.
According to ref.~\cite{DGL}, the behaviour of the T-matrix above 1.4 GeV does
not significantly affect the result. As discussed below, the solution of the
Roy equation for the S-wave is not sensitive to the asymptotics, either,
so that the contribution from low energies, which dominates the result
for the scalar radius, practically stays put. In the following, we use
the estimate given in CGL, $\langle r^2\rangle_{\mbox{\tiny
s}} = 0.61\pm0.04\,\mbox{fm}^2$. Concerning the predictions 
for the scattering lengths, the modification of the
asymptotics shifts the central values by 
\be \delta a_0^0= 0.4\times 10^{-3}\co\hspace{1em}
\delta a_0^2=0.1\times 10^{-3}\co\hspace{1em}2\,\delta a_0^0-5\,\delta
a_0^2=0.2 \times 10^{-3}\fs\ee 
In table \ref{tab:S-wave scattering lengths}, the result is compared with the
predictions of CGL.\footnote{We hope not to confuse the reader with the
  notation used in the
tables: The numbers quoted under PY are not taken from reference \cite{PY}, 
but are calculated by us, using the asymptotic
representation for the imaginary parts given there.}
Despite the fact that the error bars attached to these predictions
are very small, the above shifts amount to less than 15\% of the quoted 
uncertainties. We conclude that the values of the subtraction constants are
not affected if our asymptotics 
is replaced by the one of PY. This is of central importance, as it
confirms the statement that an accurate experimental determination of the
S-wave scattering lengths allows a crucial test of the theory.

\section{Roy equations}
\label{sec:Roy equations}
\begin{figure}[thb]
\leavevmode
\begin{center}
\includegraphics[width=10cm,angle=-90]{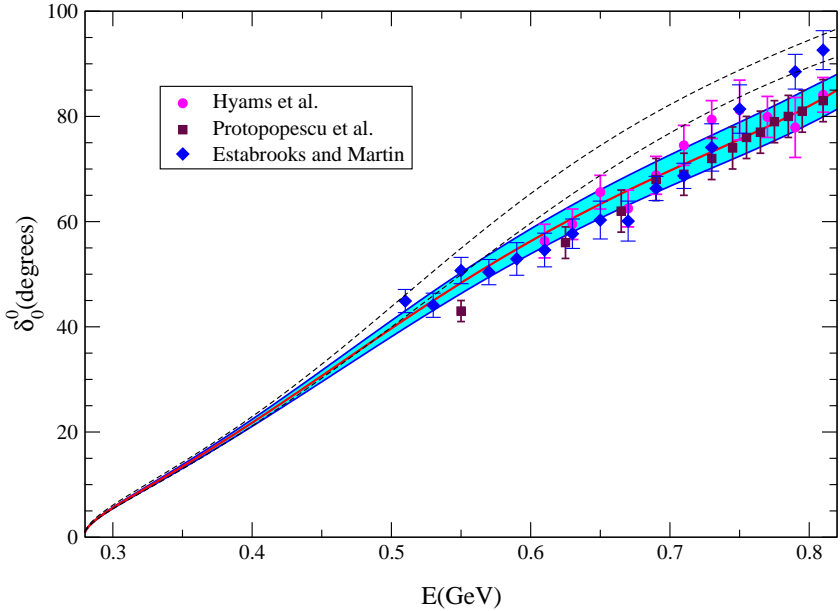}
\end{center}
\caption{\label{fig:S0}
Isoscalar S-wave.
The shaded band is taken from fig.~7 of CGL.
The full line in the middle of the band
represents the solution of the Roy equations
obtained with the asymptotics of PY.
For comparison, the representation proposed in eq.~(5.4a) of PY
is indicated by the dashed lines.}
\end{figure}
In order to determine the effect of the change in the asymptotics on
the solutions of the Roy equations, we fix the scattering lengths as well as
the phenomenological input for 0.8 GeV$\,<\,$E$\,<\,$1.42 GeV at our central
values, so that the result can be compared with our central solution. Above
1.42 GeV, we evaluate the imaginary parts with the Regge
representation of PY. The essential elements of the
calculation are described in the appendices \ref{sec:driving terms} and
\ref{sec:solution of the Roy equations}.
The result for $\delta_0^0$  is shown in
fig.~\ref{fig:S0}, 
where we compare the solution in eq.~(\ref{eq:Schenk 1})
with the band of solutions obtained in CGL.
The graph shows that the low energy behaviour of $\delta^0_0$ 
is not sensitive to the input used in the region above 1.42 GeV -- the
distortion claimed in PY does not take place. 

In PY, the ``possible cause of the distortion of the CGL solution'' is
discussed in some detail and a low energy parametrization for the isoscalar
S-wave is proposed, in support of that discussion. The proposal is referred
to as a ``tentative alternate solution'' and is indicated by the dashed lines
in fig.~\ref{fig:S0}. As can be seen from this plot, the proposal is
inconsistent both with our asymptotics and with the one of PY.
\begin{figure}[thb]
\leavevmode
\begin{center}
\includegraphics[width=12cm]{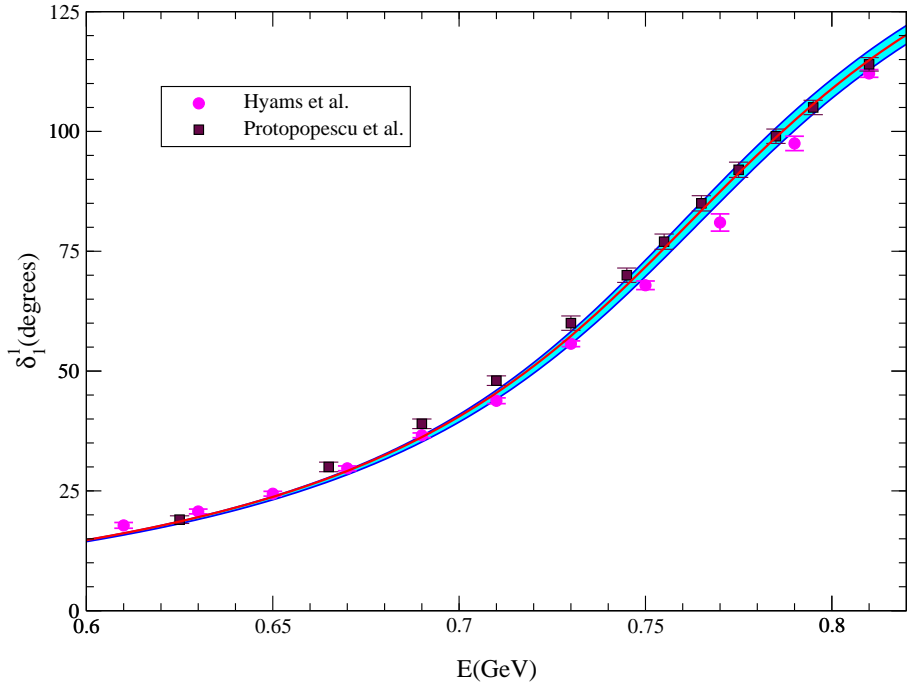}
\end{center}
\caption{\label{fig:P}
P-wave. The shaded band is taken from fig.~8 of CGL. The
full line is the solution of the Roy equations obtained with the
asymptotics of PY.}
\end{figure}

As a side remark, we note that on the interval on which we solve the Roy
equations, the various phase shift analyses are not consistent with one
another (see column 1 in table 2 of ACGL). For this reason, we did not make
use of the data on the S- and P-wave phase shifts below 0.8 GeV -- any
analysis that relies on these is subject to large uncertainties. In contrast
to the overall phase of the scattering amplitude, which is notoriously
difficult to measure, the phase difference $\delta_1^1-\delta_0^0$ shows up
directly in the cross section and is therefore known quite accurately. Indeed,
the values obtained at 0.8 GeV from the seven different phase shift analyses
listed in ACGL (which are due to Ochs \cite{ochs_phd}, Hyams et
al. \cite{hyams}, Estabrooks and Martin \cite{EM},
Protopopescu et al.~\cite{Protopopescu}, Au et al.~\cite{au} and Bugg 
et al.~\cite{bugg}) yield perfectly consistent
results for this phase difference: $\delta_1^1-\delta_0^0=26.6^\circ\pm
2.8^\circ$. While our Roy solutions agree with this experimental fact (no
wonder, we are using it in our input), the "tentative alternate 
solution" and the representation for the P-wave proposed in PY do not: These
yield  $\delta_0^0 = 91.9^\circ \pm 2.6^\circ $ and $\delta_1^1=109.0^\circ\pm
0.6^\circ$, respectively. The corresponding phase difference,
$\delta_1^1- \delta_0^0 = 17.1^\circ \pm 2.6^\circ$, is in conflict with
experiment at the level of 2.5 $\sigma$. The discrepancy must be
blamed on the "tentative alternate solution" -- the uncertainties in the
P-wave phase shift are small, because this phase is strongly constrained by
the data on the form factor (indeed the value in PY is in good agreement with
our estimate, $\delta_1^1=108.9^\circ\pm 2^\circ$). 
\begin{figure}[thb]
\centering
\includegraphics[width=12cm]{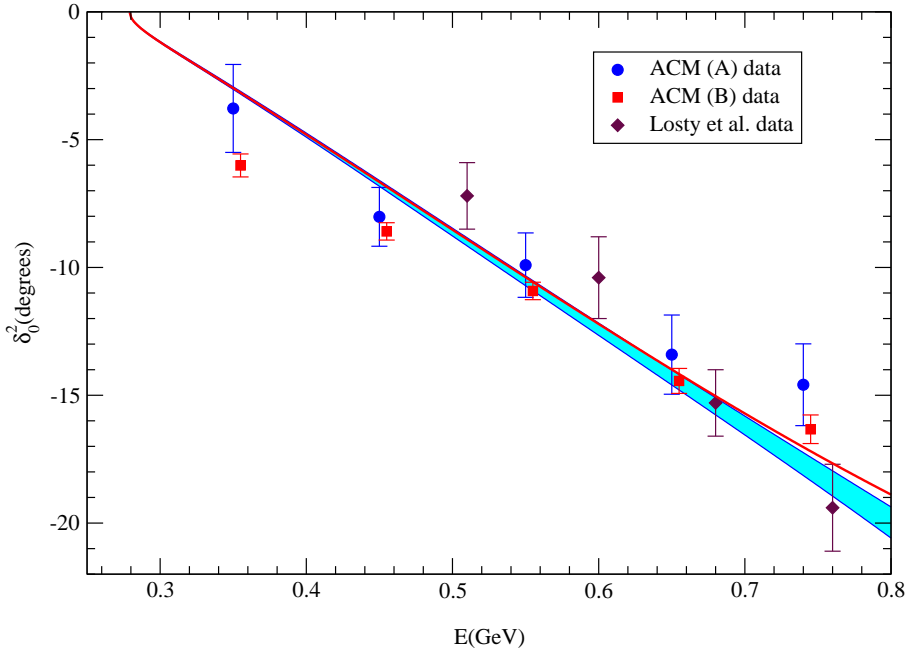}
\caption{\label{fig:S2}
Exotic S-wave. The shaded band is taken from fig.~9 of CGL. The full line
is the solution of the Roy equations obtained with the asymptotics of
PY.} 
\end{figure}

Fig.~\ref{fig:P} demonstrates that the P-wave phase shift is not sensitive to
the asymptotics, either. In the exotic S-wave (isospin 2), however, an effect
does become visible. As can be seen in fig.~\ref{fig:S2}, the modification of
the asymptotic behaviour reduces the value of $\delta_0^2$. At 0.8 GeV, the
displacement reaches $1.4^\circ$.  Although this is small compared to the
experimental uncertainties, it does imply that -- if the imaginary parts above
1.42 GeV are taken from PY -- the phase $\delta_0^2$ runs within our band of
uncertainties only below 0.64 GeV.

\section{Threshold parameters}
\label{sec:threshold parameters}Next, we evaluate
the change occurring in the result for the scattering lengths and effective
ranges of the lowest few partial waves
if our asymptotics is replaced by the one of PY. 
The evaluation is based on sum rules due to Wanders \cite{Wanders sum
rules}, which are particularly suitable here, because they are rapidly
convergent and thus not sensitive to the high energy behaviour of the imaginary
parts. The representation for $a_1^1$, for instance, reads\footnote{We use
  the normalization conventions of ref.~\cite{ACGL}. $\mbox{Im}T^I(s,0)$
  denotes the imaginary part of the forward scattering amplitude with
  $s$-channel isospin $I$. }
\bea \label{eq:a11}
a_1^1\al=\al\frac{2\,a_0^0-5\,a_0^2}{18\,M_\pi^2}+\frac{M_\pi^2}{36\,\pi^2}
\int_{4M_\pi^2}^\infty\frac{ds}{s^2(s-4M_\pi^2)^2}
 \left\{3\,(3\,s-4M_\pi^2)\,\mbox{Im}\,T^1(s,0)\right.\\
\al\al\hspace{12em} \left. 
-(s-4M_\pi^2)(2\,\mbox{Im}\,T^0(s,0)-5\,\mbox{Im}\,T^2(s,0))\right\}\fs
\nonumber\eea 
\begin{table}[H]
\begin{center}
\hspace*{-0.5em}\begin{tabular}{|c||r|r|r|}
\hline
&\multicolumn{1}{|c|}{CGL}&\multicolumn{1}{|c|}{PY}&
\multicolumn{1}{|c|}{units}\\
\hline\hspace{-0.3em}\rule{0em}{1em}
$b_0^0$&$ 0.276\pm .006$&
$\rule{0.3em}{0em}0.278$&$M_\pi^{-2}$
\\
\hline\hspace{-0.3em}\rule{0em}{1em}
$b_0^2$&$-0.803\pm .012$&$-0.800$&$10^{-1}M_\pi^{-2}$
\\
\hline\hspace{-0.3em}\rule{0em}{1em}
$a_1^1$&$0.379\pm .005$&$0.381$&$10^{-1}
M_\pi^{-2}$
\\\hline\hspace{-0.3em}\rule{0em}{1em}
$b_1^1$&$0.567\pm .013$&$0.579$&$10^{-2}
M_\pi^{-4}$
\\\hline
\end{tabular}
\end{center}
\caption{\label{tab:Wanders sum rules}
Wanders sum rules. The numbers in the first column are taken from
table 2 of CGL. Those in the second column are obtained by replacing the
asymptotics used there with the one proposed in PY.} 
\end{table}
The analogous sum rules for the effective ranges of the S- and P-waves are
listed in appendix \ref{sec:representations for threshold parameters}.
The numerical results of CGL are quoted in the first column of table
\ref{tab:Wanders sum rules}, 
while those in the second column are obtained by
repeating the calculation for the asymptotics proposed in PY.
Note that the subtraction constants play a crucial role here. In fact, $a_1^1$
is totally dominated by the contribution from the first term on the right hand
side of eq.~(\ref{eq:a11}), which accounts for 97 \% of the numerical result. 
This is why the uncertainty in our prediction for $a_1^1$ is so small. The
subtractions ensure that the integrals converge rapidly. For the asymptotics
of PY, for instance, the contributions from the region above 1.42 GeV amount
to less than 5\% of the total, for all of the quantities listed in the
table. 
\begin{table}[thb]
\begin{center}
\hspace*{-0.5em}\begin{tabular}{|r||r|r||r|r||l|}
\hline
&\multicolumn{2}{|c||}{Wanders}&\multicolumn{2}{c||}{Froissart-Gribov}&\\
\hline
&\multicolumn{1}{|c|}{CGL}&\multicolumn{1}{|c||}{PY}&\multicolumn{1}{|c|}{CGL}&\multicolumn{1}{|c||}{PY}
&\multicolumn{1}{|c|}{units}
\\\hline\hspace{-0.3em}\rule{0em}{1em}
$a^0_2$&$0.175\pm 0.003$
&$0.180$\g&$0.176$\g&$0.180$\g&
$10^{-2}M_\pi^{-4}$
\\\hspace{-0.3em}\rule[-0.5em]{0em}{0em}
$b_2^0$&$-0.355\pm 0.014$&$-0.347$\g&
$-0.359$\g&$-0.353$\g&$10^{-3}M_\pi^{-6}$
\\\hline\hspace{-0.3em}\rule{0em}{1em}
$a^2_2$&$0.170\pm 0.013$
&$0.177$\g&$0.172$\g&$0.182$\g&$10^{-3}M_\pi^{-4}$
\\\hspace{-0.3em}\rule[-0.5em]{0em}{0em}
$b_2^2$&$-0.326\pm 0.012$&$-0.327$\g&
$-0.329$\g&$-0.319$\g
&$10^{-3}M_\pi^{-6}$
\\\hline\hspace{-0.3em}\rule{0em}{1em}
$a_3^1$&$0.560\pm0.019$&$0.562$\g&$0.560$\g
&$0.565 $\g&$10^{-4}M_\pi^{-6}$
\\\hspace{-0.3em}\rule[-0.5em]{0em}{0em}
$b_3^1$&$-0.402\pm 0.018$&$-0.409$\g&$-0.404\g$&$-0.407$\g
&$10^{-4}M_\pi^{-8}$
\\\hline
\end{tabular}
\end{center}
\caption{\label{tab:D- and F-waves}
Threshold parameters of the D- and F-waves. The left half of the table lists
the results found with the analog of the Wanders sum rules,
while the numbers on
the right half are based on the Froissart-Gribov representation discussed in
section \ref{sec:FG for D-waves}. The results
obtained with the asymptotics of CGL and of PY are listed separately.
The first column is taken from table 2 of CGL.} 
\end{table}

Repeating the exercise for the D- and F-waves, we obtain the results listed on
the left half of table \ref{tab:D- and F-waves}. These indicate that the
change in the asymptotics generates a somewhat larger effect, but the
displacement stays below 5\% also here. In the case of $a_2^0$, the shift
corresponds to 1.5 
$\sigma$, while for the other quantities the prediction is not that sharp, so
that the shift is only a fraction of our error bar. 
In summary, we note that for none of the quantities considered in PY,
the change in the asymptotics proposed in that paper generates a
displacement by more than $1.5\,\sigma$.

There is an alternative method for evaluating the quantities listed in the
table: Instead of working with the analog of the Wanders sum rules, we may
invoke the Froissart-Gribov representation for the scattering lengths and
effective ranges. The difference between the two is discussed in some detail
in appendix \ref{sec:representations for threshold parameters}. If the
scattering amplitude were exactly crossing symmetric, the two methods of
calculation would yield identical results. The numerical results obtained with
the FG-representation for the P- and D-waves are discussed in sections
\ref{sec:FG for P-wave} and \ref{sec:FG for D-waves}, respectively.

The entries in columns 1 and 3 show
that, for our asymptotics, the two sets of numbers indeed agree within a
fraction of
a percent, indicating that our representation of the scattering amplitude
does pass this test of crossing symmetry. The comparison of
columns 2 and 4 indicates that the asymptotics of PY generates a somewhat
stronger violation of crossing symmetry, but the differences do not stick out
of the uncertainties that must be attached to the central values listed. As
will be discussed in section 9, however, these differences originate in the
tiny contributions from the asymptotic region and from the higher partial
waves. In fact, the slight mismatch seen in the comparison of columns 2 and 4
implies that the asymptotics of PY is not consistent with crossing symmetry.

\section{Olsson sum rule}
\label{sec:Olsson sum rule}
We now turn to the calculations described in PY
and start with the Olsson sum rule,
\bea\label{eq:Olsson} \al\al 2\,a_0^0-5\,a_0^2= O\co\eea
which relates a combination of $S$-wave scattering lengths to an
integral over the imaginary part of the forward scattering amplitude:
\be\label{eq:Olsson1} O= \frac{M_\pi^2}{8\pi^2}\int_{4M_\pi^2}^\infty\!ds\;
\frac{2\,\mbox{Im}\,T^0(s,0)+3\,\mbox{Im}\,T^1(s,0)-5\,
\mbox{Im}\,T^2(s,0)}{s\,(s-4M_\pi^2)}\,.\ee
It is well known that this integral converges only slowly --
in contrast to the subtracted dispersion integrals that underly the 
Roy equations or the sum rule for the P-wave scattering
length considered above, the contributions from the asymptotic
region play a significant role here.

In ACGL, we evaluated the integral for arbitrary values of the S-wave
scattering lengths. Inserting the predictions obtained on the basis of chiral
symmetry in eq.~(11.2) of that paper and accounting for the
correlation between $a_0^0$ and $a_0^2$  with table 4 of CGL, we obtain
$O_{\mbox{\tiny CGL}}=0.665\pm 0.022$. Since this is in perfect agreement with
our prediction for the scattering lengths, $2a_0^0-5a_0^2=0.663\pm 0.007$, we
conclude that, for our asymptotics, the Olsson sum rule is in equilibrium.

Pel\'aez and Yndur\'ain point out that if our asymptotics is replaced by
theirs, while the behaviour below 0.82 GeV is left unchanged, the value of
the integral is reduced to $O_{\mbox{\tiny PY}}=0.635\pm0.014$, so that
the sum rule gets out of equilibrium. The low energy part of their
calculation is examined in appendix \ref{sec:numerics Olsson}, where we
essentially confirm their result. Using their numbers for the
contributions from the region above 0.82 GeV, we find that the
difference $\Delta=2 a_0^0- 5 a_0^2 -O$ between the left and right hand
sides of the sum rule becomes $\Delta = 0.025\pm 0.013$, a discrepancy
of about $2\,\sigma$ (see the detailed discussion in appendix
\ref{sec:numerics Olsson}).

The result implies that the following three statements are incompatible:  
(i) the behaviour of the phases below 0.8 GeV is correctly described by the
figures shown above, (ii) the contributions above 0.8 GeV are correctly
estimated in PY, (iii) the theoretical prediction for $2a_0^0-5a_0^2$ is
valid. In PY, the blame is put on (i). The Roy equation analysis described in
section \ref{sec:Roy equations}, however, shows that (i) can only fail if
either (ii) or (iii) or both are incorrect as well. Since the phenomenological 
information leaves little room for modifications in the interval from 0.8 to
1.4 GeV, we conclude that the asymptotics proposed in PY is in conflict with
the theoretical predictions for the S-wave scattering lengths.

\section{Froissart-Gribov formula for the P-wave}
\label{sec:FG for P-wave}
In this section, we consider the Froissart-Gribov formula
\bea
a_1^1=\frac{1}{144\pi^2}\!\int_{4M_\pi^2}^\infty\frac{ds}{s^2}
\left\{2\mbox{Im}\,
T^0(s,4M_\pi^2)+3\mbox{Im}\,T^1(s,4M_\pi^2)-5\mbox{Im}\,T^2(s,4M_\pi^2)
\right\}\co \eea
which is used in PY to evaluate the P-wave scattering length.
The main difference to the Wanders representation
in eq.~(\ref{eq:a11}) is that the FG formula
does not contain a subtraction term and therefore converges
more slowly: While in the above formula, the region above 1.42 GeV is
responsible for more than 20\% of the total, only a fraction of
a percent arises from there in the case of the Wanders sum rule. The contrast
is even more pronounced in the case of $b_1^1$, where the low energy
contributions nearly cancel, so that the result obtained on the basis of the
FG formula is dominated by those from high energies:
For the asymptotics of PY, 98\% (78\%) of the total come from the region above
1.42 GeV (2 GeV). For this reason, the values found on the
basis of the FG representation come with a large
uncertainty. A numerical evaluation is of interest because it offers a test of
the input used in the asymptotic region, but 
it does not add anything of significance to our knowledge of the values of
$a_1^1$ and $b_1^1$. This is why, in table \ref{tab:Wanders sum rules},
we did not list the numerical values obtained in this way.

As both representations for $a_1^1$ are exact, the difference amounts to a
sum rule, which the imaginary parts of the scattering amplitude must
obey. Indeed, the integrand of the above representation is very similar to the
one occurring in the Olsson sum rule (\ref{eq:Olsson1}) and the
dominating contribution to the difference between the two
representations is proportional to this sum rule. The remainder 
involves the sum rule derived in appendix C of ACGL. As shown there, the
absence of a Pomeron contribution to the $I_t=1$ amplitude and crossing
symmetry imply that the integral\footnote{The barred quantities stand for 
$\mbox{Im}\,\bar{T}^I(s,t)=
\{\mbox{Im}\,T^I(s,t)-\mbox{Im}\,T^I(s,0)\}/t$. For
spacelike values of $t$, the denominator $s+t-4M_\pi^2$ develops a zero in
the range of integration, but one readily checks that the numerator $f(s,t)$
vanishes there, on account of crossing symmetry with respect to
$s\leftrightarrow u$. The same remark applies to the apparent singularity
generated by the denominator $s'-u_0$, which occurs in the
fixed-$t$ dispersion relation (\ref{eq:sub2}).}
\bea\label{eq:S(t)} \al\al
S(t)\equiv
\int_{4M_\pi^2}^\infty\!\frac{ds\,f(s,t)}{(s+t-4M_\pi^2)}\\
\al\al \rule{0em}{2.2em}f(s,t)=
\frac{2\,\mbox{Im}\,\bar{T}^0(s,t)+3\,\mbox{Im}\,\bar{T}^1(s,t)-5\,
\mbox{Im}\,\bar{T}^2(s,t)}{12\,s}-
\frac{(s-2M_\pi^2)\,\mbox{Im}\,T^1(s,0)}
{s\,(s-4M_\pi^2)\,(s - t)}\nonumber\eea
must vanish in the entire
region where the fixed-$t$ dispersion relations are valid.
Crossing symmetry
does not impose a constraint on the imaginary parts of the S-waves -- indeed,
these drop out on the right hand side of eq.~(\ref{eq:S(t)}). Hence the sum
rule $S(t)=0$ relates a family of
integrals over the imaginary part of the P-wave to the higher partial
waves. The difference between the
Froissart-Gribov and Wanders representations for $a_1^1$ may be written as
a linear combination of the Olsson sum rule and the value of $S(t)$ at
$t=4M_\pi^2$:
\bea\label{eq:a11FGW} \al\al a_1^1\,
\rule[-0.5em]{0.02em}{1.3em}_{\,\mbox{\tiny FG}}-
a_1^1\,\rule[-0.5em]{0.02em}{1.3em}_{\,\mbox{\tiny W}}=
-\frac{1}{18M_\pi^2}(2\,a_0^0-
5\,a_0^2-O)+\frac{M_\pi^2}{3\pi^2}\,S(4M_\pi^2)\fs\eea
There is an analogous formula also for $b_1^1$:
\be  b_1^1\,\rule[-0.5em]{0.02em}{1.3em}_{\,\mbox{\tiny FG}}- 
b_1^1\,\rule[-0.5em]{0.02em}{1.3em}_{\,\mbox{\tiny W}}=
\frac{1}{3\pi^2}\,\frac{\partial}{\partial t}\{t\,S(t)\}
\,\rule[-1em]{0.02em}{2.0em}_{\;t\rightarrow 4M_\pi^2}\fs\ee
Note that this relation involves the derivative with respect to $t$, because
the FG representation for $b_1^1$
contains the imaginary part as well as the first derivative thereof 
(see appendix \ref{sec:representations for threshold parameters}).

The difference between the FG and W representations for
$a_1^1$ and $b_1^1$ reflects the fact that the former is derived from an
unsubtracted dispersion relation, while the latter is based on the standard,
subtracted form. If we wish, we may just as well apply the FG
projection to the standard form of the fixed-$t$ dispersion relations.
The procedure leads to a representation that also
holds for the S-waves. In fact, the resulting formulae for $b_0^0$, $b_0^2$,
$a_1^1$ and $b_1^1$ coincide with the Wanders sum rules. In this sense,
the difference between the Froissart-Gribov and Wanders representations for
the quantities considered above exclusively concerns the manner in which the
contributions from the subtractions are dealt with. For the threshold
parameters of the higher waves, on the other hand, the subtractions do not
make any difference.

In the units of table \ref{tab:Wanders sum rules},
the numerical evaluation of the integrals yields
 \bea \al\al a_1^1\,\rule[-0.5em]{0.02em}{1.3em}_{\,\mbox{\tiny
    FG}}\simeq 0.37\hspace{2em}
b_1^1\,\rule[-0.5em]{0.02em}{1.3em}_{\,\mbox{\tiny
    FG}}\simeq 0.37\hspace{2em}\mbox{(CGL)}\co \\
\al\al a_1^1\,\rule[-0.5em]{0.02em}{1.3em}_{\,\mbox{\tiny
    FG}}\simeq 0.36\hspace{2em}
b_1^1\,\rule[-0.5em]{0.02em}{1.3em}_{\,\mbox{\tiny
    FG}}\simeq 0.56\hspace{2em}\mbox{(PY)}\fs\nonumber\eea
The second line confirms the central values given in PY:
$a_1^1\,\rule[-0.5em]{0.02em}{1.3em}_{\,\mbox{\tiny
    FG}}= 0.371\pm 0.013$, $b_1^1\,\rule[-0.5em]{0.02em}{1.3em}_{\,\mbox{\tiny
    FG}} = 0.599\pm 0.088$ (the term ``direct'' used in that paper refers
to the results obtained on the basis of the Wanders sum rules).

The above numbers show that, irrespective of the asymptotic input used, the
estimates obtained for $a_1^1$ on the basis of the FG formula are in
reasonable agreement with the much more precise result found with the
Wanders sum rule (see table \ref{tab:Wanders sum rules}). The number
extracted from the FG representation for $b_1^1$, however, is reasonably
close to the truth only for the asymptotics of PY. 

As mentioned above, the FG integral for $b_1^1$ is dominated by the
contributions from high energies. More precisely, the Regge term with the
quantum numbers of the $\rho$ is relevant, for which we are using a
parametrization of the form  $\beta_\rho(t) s^{\alpha_0+\alpha_1 t}$. 
For the integrals discussed in CGL, the uncertainty in the contribution from
this term is governed by the one in the residue $\beta_\rho(t)$, but this is
not the case here: Since the FG integral for $b_1^1$ converges only very
slowly, it is very sensitive also to the
parameters that describe the trajectory. While the values $\alpha_0 =
\frac{1}{2}-\alpha_1\,M_\pi^2 \simeq 0.49$, $\alpha_1=\frac{1}{2}
(M_\rho^2-M_\pi^2)^{-1}\simeq 0.87\,\mbox{GeV}^{-2}$ used
in CGL are determined by $M_\rho$ and $M_\pi$,  those in PY,  $\alpha_0 =
0.52\pm 0.02$, $\alpha_1 = 1.01\,\mbox{GeV}^{-2}$, are 
based on fits to cross sections. In the case of the Wanders representation for
$b_1^1$, the change occurring if the trajectory used in CGL is replaced by
the one in PY is small compared to the error in our result,
$b_1^1=0.567\pm0.013$, but in the case of the 
Froissart-Gribov representation, the operation shifts the outcome from 0.37 to
0.45 $(\alpha_0=0.50)$ or 0.53 ($\alpha_0=0.54$). Furthermore, the uncertainty
attached to the Regge residue in CGL affects the result at the level of $\pm
0.06$. Note also that the FG-representation involves 
a derivative with respect to $t$, so that not only the slope $\alpha_1$ of the
trajectory enters, but also the slope of the residue $\beta_\rho(t)$. Both of
these quantities are poorly known. 

We conclude that the FG-value for $b_1^1$ obtained with our
asymptotics is subject to a large uncertainty, because it depends on minute
details of the Regge representation: The discrepancy with the
value found in CGL is in the noise. We repeat that the issue
does not touch our prediction for $b_1^1$, for two reasons: (i) That
prediction relies on the rapidly convergent representation of Wanders, where
the entire region above 2 GeV contributes less than 2\% of the total. (ii) The
Wanders sum rule for $b_1^1$ does not involve derivatives with respect to
$t$. 

\section{Values of $a_1^1$ and $b_1^1$ from $e^+e^-$ and
    $\tau$ data} 
The data on the pion form factor can be used to arrive at an independent
determination of the P-wave parameters. As pointed out in PY, the numbers for
$b_1^1$ obtained from fits based on the method of de Troc\'oniz and
Yndur\'ain \cite{Troconiz Yndurain} disagree with our prediction at the
$4\,\sigma$ level.   
 
The partial wave parametrization used in ref.~\cite{Troconiz
Yndurain} is of inverse amplitude type:
\bea\label{eq:TY} t_1^1(s)\al=\al \left\{4\,\frac{M_\rho^2-s}{s-4M_\pi^2}\;
(B_0+B_1\, z )- i\, \sqrt{1-\frac{4M_\pi^2}{s}}\right\}^{-1}\co\no
z\al=\al \frac{\sqrt{s}-\sqrt{s_1-s}}{\sqrt{s}+\sqrt{s_1-s}}\fs
\eea 
On the interval $4M_\pi^2<s<s_1$, the square roots are real, so that the
expression obeys elastic unitarity. For $B_1 = 0$, the formula reduces to
$\rho$--meson dominance. At $s=s_1$, the term $B_1\,z$ develops a branch
cut that mimics contributions from inelastic channels. In PY, the value of
$\sqrt{s_1}$ is fixed at 1.05 GeV, while $M_\rho$, $B_0$ and $B_1$ are treated
as free parameters. Two of these specify the mass and the width of the
$\rho$, while the third describes the behaviour near threshold, which is
governed by the scattering length $a_1^1$. The authors use the above
representation to evaluate the Omn\`es factor, which accounts for the
branch cut singularity generated by the final state interaction. The remaining
singularities, in particular also the branch cuts associated with inelastic
channels, are parametrized in terms of a polynomial in a conformal
variable that is adapted to the analytic structure of the form factor. The
authors then make a fit to the $e^+e^-$ and $\tau$ data for energies 
below 0.96 GeV and come up with remarkably accurate values for the parameters
$M_\rho$, $B_0$ and $B_1$. The corresponding result for the phase at 0.8 GeV is
$\delta_1^1=109.0^\circ\pm 0.6^\circ$, while for scattering length and
effective range, the threshold expansion of the above formula yields
$a_1^1=(38.6\pm 1.2)\cdot 10^{-3}$ and $b_1^1=(4.47\pm 0.29)\cdot 10^{-3}$,
respectively. Table \ref{tab:Wanders sum rules} shows that  
the result for $a_1^1$ is consistent with our prediction, but the one for
$b_1^1$ is not. It is evident from this table that the discrepancy cannot be
blamed on the input used in the asymptotic region.

The problem with the above determination of $b_1^1$ is that it depends on the
specific form of the parametrization used for the phase. To explicitly
demonstrate this model--dependence, it suffices to allow for additional
terms in the conformal polynomial, replacing $B_0+B_1 z$ by $B_0+B_1 z+ B_2
z^2+ B_3 z^3$. For simplicity, let us fix $a_1^1$ as well as $ b_1^1$ at the
central values obtained in CGL. We can choose the remaining three parameters
in such a way that the phase stays close to the one specified 
in eq.~(3.5) of PY. With $M_\rho =773.6 \,\mbox{GeV}$,
$B_0=1.073$, $B_1=0.214$, $B_2=-0.039$, $B_3=-0.267$, for instance, the
scattering length as well as the effective range agree with the central values
in CGL and the phase stays well within the uncertainty band that follows from
the errors attached to the parameters in PY (despite the fact that these
cannot be taken literally -- above 0.82 GeV, the corresponding uncertainty in
the phase is less than half a degree). Since the available experimental
information does not strongly constrain the behaviour of the form factor in
the threshold region, it is not possible to distinguish the two
representations for the phase shift on phenomenological grounds.  

Incidentally, one may also attempt to solve the Roy equations using the
para\-metrization in eq.~(\ref{eq:TY}). The result is the same: Three
parameters do not suffice to obtain solutions that obey the Roy equation for
the P-wave, but with the above extension, the problem disappears. We
conclude that the claimed 4 $\sigma$ discrepancy is a property of the model
used to parametrize the P-wave phase shift and does not occur if one
allows for the number of degrees of freedom necessary to trust the
representation, not only for $a_1^1$, but also for $b_1^1$.
  
In connection with the contribution from hadronic vacuum polarization to
the magnetic moment of the muon, we are currently performing an analysis of
the form factor that is very similar to the one in ref.~\cite{Troconiz
Yndurain}. The main difference is that we do not invoke a parametrization
in terms of a modified Breit-Wigner formula to describe the behaviour of
the P-wave in the low energy region, but instead rely on the CGL phase
shift \cite{paper on form factor}. We obtain a perfect description of the
available experimental information about the form factor in this way,
including the data in the spacelike region and we have checked that this
also holds if we restrict our analysis to the data sets used in
\cite{Troconiz Yndurain}. By construction, our parametrization of the form
factor keeps the low energy parameters $a_1^1$ and $b_1^1$ fixed at the CGL
values. This confirms the conclusion reached above: The experimental
information on the form factor does not allow a model--independent
determination of $a_1^1$ and $b_1^1$ at the level of accuracy claimed in PY.

\section{Froissart-Gribov formula for the D-waves}
\label{sec:FG for D-waves}
Finally, we comment on the estimates for the threshold parameters of the
D-waves given in PY. Using the Froissart--Gribov representation in
eq.~(\ref{eq:FG representation}), 
the authors arrive at values for the
combinations $a_{0+}=\frac{2}{3}(a_2^0-a_2^2)$ and 
$a_{00}=\frac{2}{3}(a_2^0+2\,a_2^2)$
that differ from those obtained with the results of CGL by about 4\%. They
then argue that the two evaluations are correlated and
come up with the conclusion that, if the correlations are accounted for,
this difference amounts to a discrepancy
of 4 $\sigma$ in the case of $a_{0+}$ and 5 $\sigma$ in the
case of $a_{00}$.

The main observation concerning the comparison is that it does not allow
one to draw any conclusions about the S- and P-waves, because the
contributions from these waves are identical in the two
evaluations\footnote{ An explicit expression for the difference $\Delta
a^0_2= a_2^0\, \rule[-0.5em]{0.02em}{1.3em}_{\,\mbox{\tiny FG}}-
a_2^0\,\rule[-0.5em]{0.02em}{1.3em}_{\,\mbox{\tiny W}}$ is given in
appendix D. For all of the quantities listed in table 3, the contributions
from the S- and P-waves to the Froissart-Gribov and Wanders representations
are identical. This also holds for the $\ell = 4$ scattering lengths, but
not for the corresponding effective ranges, nor for the threshold
parameters belonging to $\ell>4$.}. When the correlations are accounted
for, these contributions drop out in the comparison. For this reason, the
differences discussed in PY between the ``CGL--direct'' and their own
calculation of the D-wave threshold parameters do not involve any of the
results of CGL. Even if the discrepancies obtained in PY could be taken at
face value, the only conclusion we could draw from these is that the Regge
representation proposed in PY differs from the one used in CGL -- but
this is evident ab initio. The situation for the Olsson sum rule and the
P-wave threshold parameters is different: In those cases, the contributions
from the S- and P-waves do not drop out, so that the CGL analysis does
enter the comparison.

Moreover, for the D-wave threshold parameters, PY do not refer to (and we
are not aware of) a determination that is independent of the sum rules. If
this were available, one could use it, together with the assumption that
the asymptotics of PY is correct, to draw conclusions about the size of the
contributions from the low-energy region and decide whether the results
obtained in CGL are consistent with those conclusions. This is the logic
the authors follow with the Olsson sum rule, where they can use the
low--energy theorem, and with the P-wave threshold parameters, where PY
claim that fits to the data on the form factor allow one to determine
$a_1^1$ and $b_1^1$ more reliably than with the sum rule (in section 8 we
explained why this is not the case). For the D-wave threshold parameters,
however, an analogous claim is not made. Hence the comparison of the
results obtained by inserting the two different representations for the
asymptotic region and for the higher partial waves in the integrals for the
threshold parameters cannot possibly lead to conclusions that go beyond the
fact that those two representations are different.

The FG integral converges almost as rapidly as the Wanders representation --
the region above 1.42 GeV only contributes a small fraction of the total. For
this reason, table 3 also lists the results obtained with the Froissart-Gribov
representation. The requirement that the two different expressions for the D-
and F-waves must lead to the same 
result amounts to a set of sum rules, which exclusively involve the
imaginary parts of the higher partial waves. The prototype of this category
of sum rules is the one in eq.~(B.7) of ACGL. Since a crossing symmetric
scattering amplitude automatically obeys these relations, they amount to a
test of crossing symmetry. The sum rules require the contributions
from the region below 1.42 GeV to be in balance with those from higher
energies. Since the low energy part is dominated by the experimentally well
determined isoscalar D-wave, the sum rules amount to a test of the
representation for the imaginary parts used in the asymptotic region (which
contain $I_t=0$ contributions from the Pomeron and $f$ poles, as well as 
poorly known terms with $I_t=2$). In 
application to PY, this test does not involve anything beyond the
parametrizations proposed in that reference for the partial waves with
$\ell\geq2$ and for the asymptotic region. For the central values of the
parameters, the difference between the results obtained from the
Froissart-Gribov and Wanders representations becomes (in the normalization
used for the D-waves in PY: scattering lengths in units of
$10^{-4}\,M_\pi^{-4}$, effective ranges in units of $10^{-4}\,M_\pi^{-6}$):
\begin{eqnarray}
\Delta a_{0+}^{\mathrm{PY}} = -0.11 \; \; , \quad \Delta
a_{00}^{\mathrm{PY}} = 0.03 \; \; , \quad \Delta b_{0+}^{\mathrm{PY}} 
= - 0.13 \; \; , \quad \Delta b_{00}^{\mathrm{PY}}=0.09 
\; \; .\nonumber
\end{eqnarray}
These numbers show that -- for the parametrizations proposed in PY -- the
two representations do not lead to the same result, so that there is an
inherent uncertainty in the values obtained for the D-wave threshold
parameters. In fact, with the exception of $a_{00}$, the above numbers are
all larger than the uncertainties quoted in PY for the discrepancy between
``CGL--direct'' and their own evaluation. Evidently, those uncertainties
are underestimated. In terms of the error attached to the comparison of
the values obtained for $b_{0+}$, for instance, their asymptotics violates
crossing symmetry at the level of 5 $\sigma$. In other words, their Regge
parametrization is not in equilibrium with the low energy structure: The
various terms from the asymptotic region roughly cancel, so that almost nothing
is left to compensate the contribution from the f$_2$(1275), which dominates
the low energy part of the integral. 

The problem does not occur with our asymptotics, for which the sum rules hold
to a remarkable degree of accuracy: with the central values of our
parameters we get
\begin{eqnarray}\Delta
  a_{0+}^{\mathrm{CGL}} = -0.006 \; \; , \quad 
\Delta a_{00}^{\mathrm{CGL}} = -0.009 \; \; , \quad 
\Delta b_{0+}^{\mathrm{CGL}} = -0.007 \; \; , \quad \Delta
  b_{00}^{\mathrm{CGL}}=-0.03\; \;.\nonumber 
\end{eqnarray}

In PY, it is stated that the discrepancies obtained with the Froissart-Gribov
formula for the effective ranges cannot be taken as seriously as those for the
scattering lengths, because the result is sensitive to the $t$-dependence of
the $I_t=2$ exchange piece. The violation of crossing
symmetry, however, also shows up in the scattering lengths: For the
parametrization proposed in PY, the net asymptotic contribution to $\Delta
a_{0+}$, for instance, is also much too small to keep the term from the
$f_2(1275)$ in balance, while for $\Delta a_{00}$, it is much too large. 

Note that the sum rules receive contributions exclusively from
(a) the Regge representation and (b) the low energy parametrization used for
the higher partial waves. Both of these contributions are small in
comparison to the net result for the threshold parameters, because that result
is dominated by the contributions from the S- and P-waves. For
the test of crossing symmetry, however, this comparison is of no significance,
because the S- and P-waves do not contribute at all. What counts is
whether or not (a) is in equilibrium with (b). 

We conclude that, while the asymptotics used in CGL is consistent with crossing
symmetry, the one proposed in PY is not. The violation is too large for the
comparison with CGL to be meaningful at the level of accuracy claimed in PY.

\section{Summary and conclusions}

The low energy analysis of the $\pi\pi$ scattering amplitude described in
CGL relies on input for the imaginary parts, which are partly taken from
experiment, partly from Regge theory. In the present paper, we have
investigated the sensitivity of the results to the input used in the
asymptotic region. The investigation is motivated by a recent paper of
Pel\'aez and Yndur\'ain, who advise the reader not to trust the results of
CGL, because in their opinion, the input used for the asymptotics is wrong.

The Regge representation of CGL is based on the work of Pennington and
Protopopescu \cite{Pennington Protopopescu} and is indeed quite different
from the one proposed in PY.  The main result of the analysis described in
the first part of the present article is that -- as far as the low energy
behaviour of the scattering amplitude is concerned -- this difference does
not matter. The input used for the imaginary parts above 1.42 GeV may be
replaced by the one advocated in PY. The outcome for the threshold
parameters of the leading partial waves remains almost the same:

\begin{itemize}
\item The predictions for the S-wave scattering lengths are practically
untouched. Expressed in terms of the uncertainty estimates given in CGL,
the changes amount to less than 0.15 $\sigma$. This is of crucial
importance, because the result implies that the subtraction constants in
the fixed-$t$ dispersion relations stay put -- the subtraction constants
are the essential parameters in the low energy domain.

\item Neither the effective ranges of the S-waves nor the threshold
parameters of the P-wave are sensitive to the input used in the asymptotic
region. The effects seen in the higher partial waves are somewhat larger,
but the only case where replacing the asymptotics of CGL by the one of PY
produces a change that exceeds our error estimates is the isoscalar D-wave
scattering length $a_2^0$, where the displacement amounts to 1.5 $\sigma$.

\item The Roy equations imply that the low energy behaviour of the
isoscalar S-wave and the P-wave remains practically unaffected by the
change in the asymptotics (see figs.~\ref{fig:S0} and \ref{fig:P}). The
exotic S-wave (isospin 2) is more sensitive, but even in that case, we find
that the phase shift at 0.8 GeV is displaced by only $1.4^\circ$ (see
fig.~\ref{fig:S2}). As witnessed by the fact that the changes in $a_0^2$
and $b_0^2$ are minute, the behaviour in the threshold region essentially
stays put also for this partial wave.

\end{itemize}
The calculation confirms the stability of our results with respect to the
uncertainties in the asymptotic region. Even if the representation proposed
in PY is assumed to be closer to the truth than the one of Pennington
and Protopopescu that we rely on, the predictions for the threshold
parameters remain essentially the same. We conclude that the
statements made by Pel\'aez and Yndur\'ain about the precision of
chiral-dispersive calculations of $\pi\pi$ scattering are incorrect.

In the second part of the present article, we have examined the
calculations described in PY. The main points to notice here are: (i) these
do not shed any light on the values of the threshold parameters and (ii) the
Regge parametrization proposed in PY cannot be valid within the 
uncertainties quoted for the parameters, because it violates
crossing symmetry. Our asymptotic representation does not have this problem. 
\begin{itemize} \item In the case of the Olsson sum rule
or the Froissart-Gribov representation for $a_1^1$, the integrals only
converge slowly, so that the result is 
sensitive to the uncertainties in the imaginary parts above 1.4 GeV. In
effect, the calculation yields a crude estimate for the combination
$2\, a_0^0-5\,a_0^2$ of subtraction constants. The comparison with the very
precise prediction obtained in ref.~\cite{CGL} shows that the
asymptotic representation proposed in PY brings the Olsson sum rule out of
equilibrium, while the one used in CGL passes the test very well. We conclude
that the Regge representation proposed in PY is not consistent with the
prediction of standard chiral perturbation theory for $2\, a_0^0-5\,a_0^2$. 

\item In the case of $b_1^1$, the Froissart-Gribov formula converges only
very slowly, so that the result is sensitive to the behaviour of the imaginary
parts at very high energies, in marked contrast to the integrals considered in
CGL, where energies above 3 GeV barely contribute. For the central parameter
values of the asymptotic representation used in CGL, the FG integral
for $b_1^1$ comes out too small. The result, however, very strongly depends on
details of the parametrization used for the Regge term with the quantum
numbers of the $\rho$, so that there is no discrepancy to speak of.

\item In PY, the method of ref.~\cite{Troconiz Yndurain} is used to arrive
at an independent determination of $a_1^1$ and $b_1^1$, based on the $e^+e^-$
and $\tau$ data. While the result for $a_1^1$ is in good agreement with our
prediction, the value for $b_1^1$ is not. We show that the uncertainties
attached to this method are underestimated. The data on
the form factor are perfectly consistent with our predictions, not only for 
$a_1^1$, but also for $b_1^1$. 

\item The Froissart-Gribov representation for the threshold parameters of the
D-waves converges about equally well as the Wanders representation used in CGL
-- in either case, the low energy region dominates. In PY, 
the difference between these two types of representation is used to test our
results for the low energy region. Actually, however, the contributions
from the S- and P-waves are identical in the two cases: A change in these
waves shifts our results by exactly the same amount as theirs. Even if the
discrepancies obtained in PY could be taken at face value, the only conclusion
we could draw from the comparison of the numbers for the threshold parameters
of the D-waves is that the asymptotic representation proposed in PY differs
from the one used in CGL -- but this is evident ab initio.  

\item The asymptotics proposed in PY violates crossing symmetry
rather strongly, while for the one used in CGL, the violations are in the
noise. In the case of $b_{0+}$, for instance, the representation of PY implies
a violation of crossing symmetry that is more than twice as large as  
the discrepancy with our result that the authors are claiming. This shows that
(i) their Regge representation cannot be valid down to 1.42 GeV and (ii) the
uncertainties are underestimated, particularly those attached to the
discrepancies obtained when comparing their results with ours.   
\end{itemize}
There is no doubt that the representation used in CGL for the asymptotic
region, as well as the one for the low energy contributions from the D- and
F-waves could be improved. In particular, the $t$-dependence of the imaginary
parts is poorly known at high energies. An improved representation could be
found by exploiting the 
various sum rules discussed in the present article and comparing the result
with what can be extracted from the experimental information about the
behaviour at high energies by invoking factorization. A better knowledge of
the imaginary parts in the region above 0.8 GeV is of interest, for
instance, in connection with the Standard Model prediction for the magnetic
moment of the muon: Our investigation of the pion form factor \cite{paper
on form factor} relies on an extension of the Roy equation analysis to
higher energies, where the uncertainties in the asymptotic region are not
entirely negligible.  Concerning the behaviour in the threshold region,
however, we do not expect this investigation to add much to what is known
already.

\section*{Acknowledgement}We thank Jos\'e Pel\'aez and Paco
Yndur\'ain for sending us the manuscript prior to publication. The present work
was carried out while one of us (H.~L.) stayed at DESY Zeuthen. He thanks
Fred Jegerlehner for a very pleasant 
stay. This work was supported by the Humboldt Foundation, by the Swiss
National Science Foundation, by SCOPES (Contract 7 IP 62607) and by RTN,
BBW-Contract No. 01.0357 and EC-Contract  HPRN--CT2002--00311 (EURIDICE).

\appendix
\renewcommand{\theequation}{\thesection.\arabic{equation}}
\setcounter{equation}{0}
\section{Asymptotic representation of PY}
\label{sec:Regge PY}
In the notation of ACGL, the Regge representation used in PY for the
imaginary parts above 1.42 GeV reads
\bea \mbox{Im}\,T^0(s,t)\al=\al \mbox{$\frac{1}{3}$}f_0(s,t) +
f_1(s,t)+\mbox{$\frac{5}{3}$}\,f_2(s,t)+
(t\leftrightarrow u)\co\no 
\mbox{Im}\,T^1(s,t)\al=\al \mbox{$\frac{1}{3}$}f_0(s,t) +
\mbox{$\frac{1}{2}$}\,f_1(s,t)-\mbox{$\frac{5}{6}$}\,
f_2(s,t)-(t\leftrightarrow u)\co\\
\mbox{Im}\,T^2(s,t)\al=\al \mbox{$\frac{1}{3}$}f_0(s,t) -\mbox{$\frac{1}{2}$}\,
f_1(s,t)+\mbox{$\frac{1}{6}$}\,f_2(s,t)+(t\leftrightarrow
u)\fs\nonumber\eea
The functions occurring here are given by
\bea
f_0(s,t)\al=\al n\;\sigma_P\,\left\{1+ k_1\,
  (\hat{s}/s)^{1/2}\right\}\, e^{b\,t}\,(s/\hat{s})^{\alpha_P(t)}\co\no
f_1(s,t)\al=\al n\left\{1+k_2\, (\hat{s}/s)^{1/2}\right\}\,\,f(s,t)\co\no
f_2(s,t)\al=\al  n\, k_4\,f(s,t)^2\, (\hat{s}/s)\co\\
f(s,t)\al=\al 
\,\sigma_\rho\,
\frac{1+\alpha_\rho(t)}{1+\alpha_\rho(0)}
\left\{(1+k_3)\,e^{b\,t}-k_3)\right\} (s/\hat{s})^{\alpha_\rho(t)}\co\no
\alpha_P(t)\al=\al 1+ \alpha_P^1
\,t\co\rule{0em}{1.2em}\hspace{2em}\alpha_\rho(t)=\alpha_\rho^0+
\alpha_\rho^1\,t\fs
\nonumber\eea  
The factor $n=4\,\pi^2$ accounts for the difference in normalization.
The scale is fixed at $\hat{s}=1\,\mbox{GeV}^2$ and the 
various parameters are assigned the values
\bea \alpha_P^1\al=\al 0.11\pm 0.03\,\mbox{GeV}^{-2}\co\hspace{2em}
 \alpha_\rho^0=0.52\pm 0.02\co\hspace{2em}
\alpha_\rho^1=1.01\,\mbox{GeV}^{-2}\co\no
\sigma_P\al=\al 3.0\pm 0.3\co\hspace{2em}\sigma_\rho=0.85\pm0.10\co
\hspace{2em}b=2.38\pm0.20\,\mbox{GeV}^{-2}\co\\ 
k_1\al=\al 0.24\co\hspace{1em} k_2= 0.4\pm0.1\co\hspace{1em}
k_3=1.48\co\hspace{1em}k_4=0.8\pm0.2\fs\nonumber\eea 
The value of $\sigma_P$ corresponds to an asymptotic cross section of
$n\hspace{0.07em}\sigma_P/3\hspace{0.07em}\hat{s}\simeq 15\, \mbox{mb}$. 

\section{Driving terms for asymptotics of 
PY}\setcounter{equation}{0}
\label{sec:driving terms}
The contributions to the Roy equations that arise from the imaginary parts of
the higher partial waves ($\ell\geq 2$) and from the high energy end of the
dispersion integrals are referred to as driving terms. We evaluate the former
as described in detail 
in ACGL, except that the integrals are now cut off at 1.42 GeV. Concerning the
latter, we merely have to replace the 
Regge representation used in ACGL by the one of PY and take the
lower limit of the integral over the energy at 1.42 GeV instead of 2 GeV.
The result is well approximated by polynomials in
$q^2=\frac{1}{4}(s-4M_\pi^2)$: 
\bea\label{dPY}
d^0_0(s)_{\mbox{\tiny PY}}\al =\al 0.16\,q^2 + 5.5\,q^4 -
     6.6\,q^6 + 7.5\,q^8\co\no
d^1_1(s)_{\mbox{\tiny PY}}\al=\al 0.0011\,q^2 + 0.52\,q^4 +
     0.13\,q^6 + 1.1\,q^8
\label{dttot}\co\\
d^2_0(s)_{\mbox{\tiny PY}}\al =\al 0.074\,q^2 + 2.7\,q^4 -
     7.0\,q^6 + 8.8\,q^8\co
\nonumber\eea
where $q$ is taken in GeV units.
The corresponding Roy equations are obtained by inserting these expressions
in eqs.~(5.1), (5.2) of ACGL, with $s_0= (0.8 \,\mbox{GeV})^2$, $s_2= (1.42
\,\mbox{GeV})^2$. 

\section{Roy solution for asymptotics of PY}\setcounter{equation}{0}
\label{sec:solution of the Roy equations}
In order to study the effect of the change in the asymptotics on the solutions
of the Roy equations, we fix the subtraction constants at $a_0^0=0.22$,
$a_0^2=-0.0444$ and use our central values for the phenomenological input
below 1.42 GeV, which are characterized by $\delta_0^0(s_0)=82.0^\circ$,
$\delta_1^1(s_0)=108.9^\circ$.
We describe the phases with a parametrization of the form
\be
\tan \delta_\ell^I = \sqrt{1-{4 M_\pi^2 \over s}}\; q^{2 \ell} 
\left\{A^I_\ell
+ B^I_\ell q^2 + C^I_\ell q^4 + D^I_\ell q^6 \right\} \left({4
  M_\pi^2 - s^I_\ell \over s-s^I_\ell} \right) \;.\ee
Using the driving terms in 
eq.~(\ref{dPY}), we then obtain the following solution of the
Roy equations (here, all quantities are given in units of $M_\pi$):
\bea \label{eq:Schenk 1}
A_0^0\al=\al .220\co\hspace{4.5em}
A^1_1=.380\cdot 10^{-1}\co\hspace{2.2em}
A^2_0= -.0444\co\no
 B_0^0\al=\al .269\co\hspace{4.5em}
B^1_1=  .144\cdot 10^{-3}\co\hspace{2em}
B^2_0= -.936\cdot 10^{-1}\co\no
C_0^0\al=\al -.137\cdot 10^{-1}\co\hspace{1em}
C^1_1= -.719\cdot 10^{-4}\co\hspace{1.3em}
C^2_0= -.119\cdot 10^{-1}\co\label{RoyPY}\\
D_0^0\al=\al -.146\cdot 10^{-2}\co\hspace{1em}
D^1_1=  -.117\cdot 10^{-5}\co\hspace{1.2em}
D^2_0= .415\cdot 10^{-3}\co\no
s^0_0\al=\al 36.73\co\hspace{4.4em}s_1^1=30.70\co\hspace{4.5em}s^2_0=-8.84
\fs\nonumber\eea

\section{\hspace{-0.3em}Representations for the threshold parameters}\setcounter{equation}{0}
\label{sec:representations for threshold parameters}
\subsection*{Subtractions}\setcounter{equation}{0}
As discussed in the text, the subtractions play a central role in the low
energy analysis. The fixed-$t$ dispersion relations are
needed in order to derive the various representations for the threshold
parameters used in the text. We first write these relations down explicitly.

If the subtractions are ignored, the fixed-$t$ dispersion relations are very
simple: 
\be\label{eq:sub1}\vec{T}(s,t)=\int_{4M_\pi^2}^\infty ds'
\left\{\frac{\mbox{Im}\,
\vec{T}(s',t)}
{(s'-s)}+\frac{C_{su}\cdot\mbox{Im}\,\vec{T}(s',t)}
{(s'-u)}\right\}\co\ee
where $\vec{T}(s,t)=\{\vec{T}^0(s,t),\vec{T}^1(s,t),\vec{T}^2(s,t)\}$ 
is the vector formed with the three $s$-channel isospin
components and $C_{su}$ is the $3\times 3$ crossing matrix relevant for
$s\leftrightarrow u$. The dispersion integral diverges, however.
In order to remove the divergent piece, a subtraction term of the form
$\vec{c}_0(t) +s\,\vec{c}_1(t)$ is needed.
As shown by Roy \cite{Roy}, crossing symmetry implies that the subtraction
functions $\vec{c}_0(t)$ and  $\vec{c}_1(t)$ are fully
determined by the imaginary parts of the forward scattering 
amplitude, except for two constants. The dispersion relations then take the
form 
\bea
\label{eq:sub2} 
\vec{T }(s,t)\al=\al(4M_\pi^2)^{-1}\,(s\,{\bf 1} + t\,
C_{st} + u\, C_{su})\cdot \vec{T}(4M_\pi^2,0) \\
\al\al+\int_{4M_\pi^2}^\infty
\!ds'\,g_2(s,t,s')\cdot\mbox{Im}\,\vec{T}(s',0)
+\int_{4M_\pi^2}^\infty \!ds'\,g_3(s,t,s')\cdot\mbox{Im}\,\vec{T}(s',t)
\,. \nonumber
\eea
The first term is fixed by the $S$-wave scattering lengths: 
\bea
\vec{T}(4M_\pi^2,0)=32\,\pi\,(a_0^0,0,a_0^2)\nonumber\,.
\eea
The quantities $g_2$ and $g_3$ are built with the
crossing matrices $C_{st}$, $C_{tu}$ and $C_{su}$:
\bea\label{eq:g2g3} g_2(s,t,s')&=&-\frac{t}{\pi\, s'\,(s'-4M_\pi^2)}\,
(u\, C_{st} + s\, C_{st}\, C_{tu})\left(\frac{{\bf 1}}{s'-t}
+ \frac{C_{su}}{s'-u_0}\right)\co\no
g_3(s,t,s')&=&-\frac{s\,u}{\pi\,s'(s'-u_0)}\left(\frac{{\bf 1}}{s'-s}
+ \frac{C_{su}}{s'-u}\right)\co\eea
with $u=4M_\pi^2-s-t$ and $u_0=4M_\pi^2-t$. One readily checks that the
difference between the right hand sides of eqs.~(\ref{eq:sub1}) and
(\ref{eq:sub2}) is linear in $s$.

The scattering amplitude is invariant under the crossing operations
$s\leftrightarrow t$, $s\leftrightarrow u$  and $t\leftrightarrow u$: 
$\vec{T}(s,t)=C_{tu}\cdot\vec{T}(s,u)=C_{st}\cdot\vec{T}(t,s)=
C_{su}\cdot\vec{T}(u,t)$.
These relations impose constraints on the imaginary part of the amplitude,
which can be expressed in the form of sum rules \cite{Wanders 1969,Tryon,
Basdevant Schomblond}. In particular, inserting the dispersion relation
(\ref{eq:sub2}) on the two sides of the equation  
\be\label{eq:crossing symmetry} \vec{T}(s,t)=C_{st}\cdot\vec{T}(t,s)\co\ee
one obtains an entire family of such sum rules. Note that the relation
$S(t)=0$, which we made use of in section \ref{sec:FG for P-wave}, was
considered long ago and was exploited to study the $t$-dependence of the
residue of the Regge pole with the quantum numbers of the $\rho$
\cite{Basdevant Schomblond}. For a detailed discussion, 
we refer to \cite{Pennington}.

\subsection*{Wanders representation}
The threshold parameters are the coefficients occurring in the expansion of
the scattering amplitude around the point $s=4M_\pi^2$, $ t=0$. Setting
\bea s=4\,(M_\pi^2+q^2)\co\hspace{2em}t=-2\,q^2\,(1-z)\fs\eea
and performing the expansion in powers of $q$ in the
integrands on the
right hand side of the dispersion relation (\ref{eq:sub2}), we arrive at
the Wanders sum rules:
\bea\label{eq:Wanders sum rules}
 b_0^0\al=\al\frac{1}{3M_\pi^2}(2a_0^0-5a_0^2)+\frac{M_\pi^2}{6\pi^2}
\int_{4M_\pi^2}^\infty\frac{ds\, B_0^0(s)}{s^2(s-4M_\pi^2)^2}-\beta
\,(a_0^0)^2\no
b_0^2\al=\al-\frac{1}{6M_\pi^2}(2a_0^0-5a_0^2)+\frac{M_\pi^2}{12\pi^2}
\int_{4M_\pi^2}^\infty\frac{ds\, B_0^2(s)}{s^2(s-4M_\pi^2)^2}
-\beta\,(a_0^2)^2 \no
a_1^1\al=\al\frac{2\,a_0^0-5\,a_0^2}{18\,M_\pi^2}+\frac{M_\pi^2}{36\,\pi^2}
\int_{4M_\pi^2}^\infty\frac{ds\,A_1^1(s)}{s^2(s-4M_\pi^2)^2}\\
b_1^1\al=\al\frac{1}{36\pi^2}\int_{4M_\pi^2}^\infty
\frac{ds\,B^1_1(s)}{s^3(s-4M_\pi^2)^3}\nonumber\fs\eea
The integrands are given by
\bea B_0^0(s)\al=\al4(s-M_\pi^2)\mbox{Im}\,T^0(s,0)+
(s-4M_\pi^2)\{-3\mbox{Im}\,T^1(s,0)+5\mbox{Im}\,T^2(s,0)\}-\beta^0(s)\co\no
B_0^2(s)\al=\al (s-4M_\pi^2)\{2\mbox{Im}\,T^0(s,0)+
3\mbox{Im}\,T^1(s,0)\}+(7s-4M_\pi^2)\mbox{Im}\,T^2(s,0))-\beta^2(s)\co\no
A_1^1(s)\al=\al 3\,(3\,s-4M_\pi^2)\,\mbox{Im}\,T^1(s,0)
+(s-4M_\pi^2)\{-2\,\mbox{Im}\,T^0(s,0)+5\,\mbox{Im}\,T^2(s,0)\}\co
\no
B_1^1(s)\al=\al 3\,(3\,s^3-12M_\pi^2 s^2+48 M_\pi^4 s-64M_\pi^6)\,
\mbox{Im}\,T^1(s,0) \no
\al\al
+(s-4M_\pi^2)^3\{-2\,\mbox{Im}\,T^0(s,0)+5\,\mbox{Im}\,T^2(s,0)\}\fs
\nonumber\eea
For the S-wave effective ranges, the expansion can be interchanged with the
integration only after removing the threshold singularity. This can be done by
supplementing the integrand with a total derivative, which gives rise to
extra terms in the expressions for $b_0^0$ and $b_0^2$:
\bea \beta^0(s)\al=\al \frac{48\,\pi}{M_\pi^2}\,(a_0^0)^2\,h(s)
\, \theta(s_c-s)\co\hspace{2em}
\beta^2(s)=\frac{96\,\pi}{M_\pi^2}\,(a_0^2)^2\,h(s)\, \theta(s_c-s)\co\no
\beta\al=\al\frac{8}{\pi}\int_{s_c}^\infty\frac{ds}{s^2(s-4M_\pi^2)^2}\,h(s)\co
\hspace{2em}h(s)=(s-2M_\pi^2)\sqrt{s(s-4M_\pi^2)}\fs\nonumber\eea
By construction, the result is independent of $s_c$.

The corresponding representations for the threshold parameters of the higher
waves are obtained in the same manner -- we are referring to all of these as
Wanders representations. The one for the D-wave scattering lengths, 
for instance, takes the form
\bea
a^0_2\al=\al\frac{1}{90\pi^2}\int_{4M_\pi^2}^\infty
\frac{ds\,A_2^0(s)}{s^3(s-4M_\pi^2)^2} +\frac{M_\pi^2}{45\pi^2}
\int_{4M_\pi^2}\frac{ds\,\dot{A}^0_2(s)}{s^2(s-4M_\pi^2)^2}\\
 a^2_2\al=\al\frac{1}{180\pi^2}\int_{4M_\pi^2}^\infty 
\frac{ds\,A_2^2(s)}{s^3(s-4M_\pi^2)^2}
+\frac{M_\pi^2}{90\pi^2}\int_{4M_\pi^2}\frac{ds\,\dot{A}^2_2(s)}
{s^2(s-4M_\pi^2)^2}\fs\nonumber\eea
In this case, the integrands 
\bea A_2^0(s)\al=\al 3\,(s^2+4M_\pi^2s-16 M_\pi^4)\,\mbox{Im}\,
T^1(s,0)\no
\al\al
+(s-4M_\pi^2)^2\{\mbox{Im}\,T^0(s,0)+5\,\mbox{Im}\,T^2(s,0)\}\co\no
\dot{A}_2^0(s)\al=\al
4(s-M_\pi^2)\mbox{Im}\,\dot{T}^0(s,0)
+(s-4M_\pi^2)\{-3\,\mbox{Im}\,\dot{T}^1(s,0)+
5\,\mbox{Im}\dot{T}^2(s,0)\}\co \no
A_2^2(s)\al=\al -3\,(s^2+4M_\pi^2s-16 M_\pi^4)\,\mbox{Im}\,
T^1(s,0)\no
\al\al
+(s-4M_\pi^2)^2\{2\,\mbox{Im}\,T^0(s,0)+\,\mbox{Im}\,T^2(s,0)\}\co\no
\dot{A}_2^2(s)\al=\al
(s-4M_\pi^2)\{2\,\mbox{Im}\,\dot{T}^0(s,0)+3\,\mbox{Im}\,\dot{T}^1(s,0)\}\no
\al\al+(7\,s-4M_\pi^2)\,\mbox{Im}\,\dot{T}^2(s,0)
\co\nonumber\eea
involve the first derivative of the scattering
amplitude with respect to $t$,
\bdm \dot{T}^I_\ell(s,t)\equiv \frac{\partial}{\partial t} T^I_\ell(s,t)
\nonumber\edm
We do not list the analogous expressions for the D-wave effective ranges
or for the F-wave scattering length. These are obtained with the same
algorithm and involve up to two derivatives.

\subsection*{Froissart-Gribov representation}
The crossing relation (\ref{eq:crossing symmetry}) connects the properties of
the amplitude in the vicinity of threshold to those
in the vicinity of the  point $s=0$, $t=4M_\pi^2$. The Froissart-Gribov
representation of
the threshold parameters may be obtained by inserting the unsubtracted
dispersion relation (\ref{eq:sub1}) in eq.~(\ref{eq:crossing symmetry}) and
expanding the result around $s=4M_\pi^2$, $t=0$.
Instead of the Wanders sum rules, we now obtain
\bea\label{eq:FG representation}  
a_\ell^I\al=\al\frac{2^\ell\,\ell ! }{16\,\pi^2\,(2\ell+1)!!}
\;\int_{4M_\pi^2}^\infty ds\;\frac{\mbox{Im}\,
T^{(I)}(s,4M_\pi^2)}{s^{\,\ell+1}}\co\\
b_\ell^I\al=\al \frac{2^\ell\,\ell ! }{8\,\pi^2\,(2\ell+1)!!}
\;\int_{4M_\pi^2}^\infty ds\;\frac{2\,s \,\mbox{Im}\,
\dot{T}^{(I)}(s,4M_\pi^2)-(\ell+1)\mbox{Im}\,
T^{(I)}(s,4M_\pi^2)}{s^{\,\ell+2}}\nonumber \eea
The quantities $T^{(0)}(s,t)$, $T^{(1)}(s,t)$ and $T^{(2)}(s,t)$
represent the $t$-channel isospin components of the scattering amplitude,
\bea T^{(I)}(s,t)=\sum_{I'}C_{st}^{I I'}\,T^{I'}(s,t)\eea
and $\dot{T}^{(I)}(s,t)$ stands for the derivative of $T^{(I)}(s,t)$ with
respect to $t$. 

In view of the occurrence of subtractions, the representation
holds in this form only for $\ell\geq 1$. In order to arrive at a
representation that also holds for the S-waves, it suffices to insert 
in eq.~(\ref{eq:crossing symmetry}) the
subtracted version (\ref{eq:sub2}) of the dispersion relation rather than the
unsubtracted one. The
subtractions are linear in $s$. After crossing, they become linear in $t$ and
thus drop out in all waves except S and P. So the expressions for the
threshold parameters remain the same for $\ell \geq 2$. On the other hand, the
term containing the function $g_3(s,t,s')$ in eq.~(\ref{eq:g2g3})
is proportional to $s\, u$. After
crossing, this becomes $t\, u =4
\,q^4(1-z^2)$. So, the term does not contribute to the scattering lengths or
effective ranges of the S- and P-waves. Hence the resulting representation for
these exclusively contains the imaginary parts in the forward direction. 
In fact, the representation for
$b_0^0$, $b_0^2$, $a_1^1$, $b_1^1$ that obtains in this manner is identical
with the Wanders sum rules in eq.~(\ref{eq:Wanders sum rules}). The exercise
shows that the lowest terms in the threshold expansion of the subtracted
fixed-$t$ dispersion relations automatically respect crossing symmetry.

\subsection*{Sum rule related to $a_2^0$}
As mentioned in the text, the contributions from the S- and P-waves to the FG
and W representations of the D- and F-wave threshold parameters are identical.
For $\Delta a_2^0=a_2^0\,
\rule[-0.5em]{0.02em}{1.3em}_{\,\mbox{\tiny FG}}-
a_2^0\,\rule[-0.5em]{0.02em}{1.3em}_{\,\mbox{\tiny W}}$, for instance, the
explicit expression reads
\begin{eqnarray}
&&\hspace{-5em}\Delta a^0_2= \frac{16}{45\pi} \sum_{\ell} (2 \ell +1)
\int_{4 M_\pi^2}^{\infty}\frac{ds}{s^3}\,\left\{\rule[-1em]{0em}{2em}
\bar{P}_\ell(z_s)
\left\{\mathrm{Im}t^0_\ell(s)+3\, \mathrm{Im}t^1_\ell(s)+
5\, \mathrm{Im}t^2_\ell(s)\right\}\right.
\nonumber \\
&&\hspace{-3em}\left.-\, \frac{2 M_\pi^2 s}{(s-4M_\pi^2)^3} \left\{
\kappa^0_\ell(s)\, 
\mathrm{Im}t^0_\ell(s) + \kappa^1_\ell(s)\, \mathrm{Im}t^1_\ell(s) + 
\kappa^2_\ell(s)\, \mathrm{Im}t^2_\ell(s)\right\}\right\}\;,
\nonumber  \\
\hspace{3em}
z_s&\!\! =&\!\!\frac{s+4M_\pi^2}{s-4M_\pi^2}\,,\hspace{3em}
\bar{P}_\ell(z)= \left\{\begin{array}{ll}
P_\ell(z)-1 &
\mbox{$\ell$ even}\nonumber \\ 
 P_\ell(z)-z &
\mbox{$\ell$ odd}\end{array}\right.\nonumber\\
\kappa_\ell^0(s)&\!\! =&\!\! 4\,\ell(\ell+1)(s-M_\pi^2)\;,\nonumber\\
\kappa_\ell^1(s)&\!\! =&\!\!-3(\ell-1)(\ell+2)(s-4M_\pi^2)\;,\nonumber\\
\kappa_\ell^2(s)&\!\! =&\!\!5\,\ell(\ell+1)(s-4M_\pi^2)\;.\nonumber
\end{eqnarray}
In the notation used here, the sum extends over all values of $\ell$, but
$\mathrm{Im}t^0_\ell(s)$
and $\mathrm{Im}t^2_\ell(s)$ are different from zero only if $\ell$ is even,
while $\mathrm{Im}t^1_\ell(s)$ vanishes unless $\ell$ is odd. The formula
explicitly demonstrates that the S- and P-waves do not 
contribute to the sum rule $\Delta a_2^0=0$: The coefficients
$\bar{P}_\ell(z)$ and $\kappa_\ell^I(s)$ vanish for $\ell=0$ and $\ell=1$. 

\section{Numerics for the Olsson sum rule}\setcounter{equation}{0}
\label{sec:numerics Olsson}

In PY, the contributions to the Olsson integral arising from the imaginary
parts of the S-and P-waves below 0.82 GeV are
estimated at $0.400\pm0.007$. The central value is in good agreement
with what is obtained with the central solution in eq.~(17.2) of CGL:
$ O_{\mbox{\tiny SP}}(E<0.82\, \mbox{GeV})=0.401$  (no wonder: it is
calculated from this solution, except that
an  extrapolation for the interval from 0.80 to 0.82 GeV is made). 
If we instead use the Roy solution relevant for the asymptotics of PY, which
is specified in eq.~(\ref{RoyPY}), we
obtain $ O_{\mbox{\tiny SP}}(E<0.82\, \mbox{GeV})=0.404$. The comparison
demonstrates that the low energy behaviour of the integrand in the Olsson sum
rule is not sensitive to the asymptotics. Concerning the error bar to be
attached to the 
central value, we note that the uncertainties in the low energy
theorems for the S-wave scattering lengths generate an error of $\pm0.005$, 
while those in the phases at the
matching point affect the result by $\pm0.007$. The
noise in the experimental input used in the region from 0.8 to 1.42 GeV
also generates some uncertainty in the Roy solutions. We investigate this
effect by comparing the results found for the different phase shift analyses
shown in Fig.~3 of ACGL. The error from this source is small - we estimate it
at $\pm 0.002$. Finally, we take the difference between
the central solutions belonging to the asymptotics of ACGL and PY
as an estimate of the uncertainties from the region above 1.42 GeV. The
net result then reads 
\bea\label{OSP}  O_{\mbox{\tiny SP}}(E< 0.82\, \mbox{GeV})\al=\al 0.404\pm
0.005\pm 0.007\pm 0.002\pm 0.003\fs\eea
This adds up to $0.404\pm 0.009$, in good agreement with the value
$0.400\pm0.007$ obtained in PY. Adding the various terms listed in
eq.~(4.3) of PY, we obtain
\be \label{O} O_{\mbox{\tiny PY}} = 0.638\pm 0.015\fs\ee
So, the change in the asymptotics proposed in PY indeed
pulls the Olsson integral down, by about 
0.029 and thus tends to bring the sum rule out of equilibrium.

The left hand side of the Olsson sum rule is determined by the S-wave
scattering lengths. These also enter the above calculation of the right
hand side: The first error in eq.~(\ref{OSP}) reflects the uncertainties due
to this source. The remaining terms on the right hand side of this
equation as well as the contributions from $E>0.82\,\mbox{GeV}$ are
independent of $a_0^0, a_0^2$, so that the net uncertainty in the
difference between the two sides of the Olsson sum rule cannot be smaller
than the errors that remain if the uncertainty on the left as well as the
first error in eq.~(\ref{OSP}) are dropped. Indeed, the two terms mentioned
nearly cancel: Varying the S-wave scattering lengths in the error ellipse
given in CGL, the quantity $\Delta\equiv 2a_0^2-5a_0^2-O$ only varies by
$\pm0.002$. Adding the other sources of uncertainty, we obtain $\Delta =
0.025\pm 0.013$ and thus confirm the result $\Delta=0.027\pm
0.011$ quoted in PY.

\end{document}